\documentclass{PoS}

\title{$CP$ violation in charmed hadron decays into neutral kaons}

\ShortTitle{$CP$ violation in charmed hadron decays into neutral kaons}

\author{Di Wang\\
        School of Nuclear Science and Technology,  Lanzhou University,
Lanzhou 730000,  People's Republic of China\\
        E-mail: \email{dwang15@lzu.edu.cn}}

 \author{Fu-Sheng Yu\\
      School of Nuclear Science and Technology,  Lanzhou University,
Lanzhou 730000,  People's Republic of China\\
      E-mail: \email{yufsh@lzu.edu.cn}}

\author{Hsiang-nan Li\\
       Institute of Physics, Academia Sinica, Taipei, Taiwan 115, Republic of China\\
        E-mail: \email{hnli@phys.sinica.edu.tw}}

\abstract{~~~We find a new $CP$ violating effect in charmed hadron decays into neutral kaons, which is induced by the interference between the Cabibbo-favored and doubly Cabibbo-suppressed amplitudes with the $K^{0}-\overline K^{0}$ mixing \cite{Yu:2017oky}. It is estimated to be of order of $\mathcal{O}(10^{-3})$, much larger than the direct $CP$ asymmetry, but missed in the literature. To reveal this new $CP$ violation effect, we propose a new observable,
the difference of the $CP$ asymmetries in the
$D^{+}\to \pi^{+}K_S^0$ and $D_{s}^{+}\to K^{+} K_S^0$ modes.
Once the new effect is determined by experiments, the direct $CP$ asymmetry then can be extracted and used to search for new physics.
}

\FullConference{The 15th International Conference on Flavor Physics \& CP Violation\\
		 5-9 June 2017 \\
		 Prague, Czech Republic}

\usepackage{amsmath,bm}
\usepackage{braket}

\begin{document}

$CP$ asymmetry plays an unique role in understanding the matter-antimatter asymmetry and searching for new physics. It has been well established in the
kaon and $B$ meson systems \cite{Christenson:1964fg,Aubert:2001nu,Abe:2001xe,PDG},
but not yet in the charm sector. In the past decade, many efforts
have been devoted to study the $CP$ violation in the singly Cabibbo-suppressed
(SCS) $D$ meson decays. The most precise experimental result of $CP$ violation in SCS decays is \cite{Aaij:2016cfh}
\begin{align}
\Delta A_{CP} \equiv A_{CP}(D^0\to K^+K^-)- A_{CP}(D^0\to \pi^+\pi^-)=(-0.10\pm0.08\pm0.03)\%.\label{data}
\end{align}
 With the precision lower than $10^{-3}$, the $CP$ violation in charm decays has not been observed.

 $CP$ asymmetry can also occur in $D\to fK^0_S$ decays, where $f$ is a final-state particle. For example, the $CP$ violation in $D^+\to \pi^+K^0_S$ has been measured by Belle collaboration with $3.2\sigma$ from zero \cite{Ko:2012pe}. In this work,
We point out a new $CP$-violation effect, which is induced by the interference
between the Cabibbo-favored (CF) and doubly Cabibbo-suppressed (DCS) amplitudes with the mixing of final-state
mesons \cite{Yu:2017oky}. It is estimated to be of order of $10^{-3}$, much larger than the direct $CP$ asymmetry,
however, missed in the literature
\cite{Ko:2012pe,Lipkin:1999qz,Grossman:2011zk,Bianco:2003vb}.
We propose a new observable, the difference between the
$CP$ asymmetries in the $D^{+}\to \pi^{+}K^0_S$
and $D_{s}^{+}\to K^{+}K^0_S$ decays, to measure the new $CP$ violation effect. Once the new effect is obtained, the direct $CP$ asymmetry in charm decays can be extracted correctly and used to search for new physics.

In experiments, the $K^0_S$ state is reconstructed by $\pi^+\pi^-$ final state. The time-dependent $CP$ violation in $D$ meson decays into neutral kaons is defined by
\begin{equation}\label{m1}
A_{CP}(t) \equiv\frac{\Gamma(D\to K(t)(\to \pi^+\pi^- )f)-\Gamma(\overline D\to K(t)(\to \pi^+\pi^-)\overline f)}{\Gamma(D\to K(t)(\to \pi^+\pi^- )f)+\Gamma(\overline D\to K(t)(\to \pi^+\pi^-)\overline f)},
\end{equation}
where  $K(t)$ donates the immediate state of neutral kaons.
The mass eigenstates of neutral kaons, $|K_S^0\rangle$ of mass $m_S$ and width $\Gamma_S$ and $|K_L^0\rangle$ of mass $m_L$ and width $\Gamma_L$, are linear combinations of the flavor eigensates $|K^0\rangle$ and $|\overline K^0\rangle$,
$|K_{S,L}^0\rangle  =   p_K|K^0\rangle\mp q_K|\overline{K}^0\rangle$,
with $p_K =   (1+\epsilon)/\sqrt{2(1+|\epsilon|^2)}$ and
$q_K =   (1-\epsilon)/\sqrt{2(1+|\epsilon|^2)}$, and $\epsilon$ is a small parameter characterizing the indirect $CP$ violation in neutral kaon mixing \cite{PDG}.
For convenience, the ratio between DCS and CF amplitudes is set as
 \begin{equation}\label{r}
 \mathcal{A}(D\rightarrow K^0f)/\mathcal{A}(D\rightarrow \overline{K}^0f) = r_f\,e^{i(\phi+\delta_f)},
 \end{equation}
where $r_f$ is the size of the ratio, $\phi$ and $\delta_f$ are relative weak and strong phases respectively. In the SM,
$r_f\sim|V_{cd}^{*}V_{us}/V_{cs}^{*}V_{ud}|\sim\mathcal{O}(10^{-2})$ and
$\phi\equiv Arg\left[-V_{cd}^{*}V_{us}/V_{cs}^{*}V_{ud} \right] =(-6.17\pm 0.43)\times 10^{-4}$ \cite{PDG}.
With the small parameters $\epsilon$, $r_f$ and $\phi$, we obtain the time-dependent $CP$ violation as
\begin{equation}\label{eq:KSAcp}
A_{CP}(t)\simeq\big[A_{CP}^{\overline K^0}(t)+A_{CP}^{\rm dir}(t)+A_{CP}^{\rm int}(t)\big]/D(t),
\end{equation}
in which
 $D(t)= e^{-\Gamma_St}(1-2r_f\cos\delta_f\cos\phi)+e^{-\Gamma_Lt}|\epsilon|^2$.
The $A_{CP}^{\overline K^0}(t)$ term is the indirect $CP$ violation in $K^0-\overline K^0$ mixing,
\begin{align}\label{po1}
A_{CP}^{\overline K^0}(t)
=  -2e^{-\Gamma t}
\big(Re(\epsilon)\cos(\Delta mt)+Im(\epsilon)\sin(\Delta mt)\big)+2Re(\epsilon)e^{-\Gamma_St}.
\end{align}
The $A_{CP}^{\rm dir}(t)$ term is the direct $CP$ violation in charm decay induced by the interference between the CF and DCS amplitudes,
\begin{align}
A_{CP}^{\rm dir}(t)&=2e^{-\Gamma_St}\,r_f\sin\delta_f\sin\phi.
 \end{align}
The $A_{CP}^{\rm int}(t)$ term is the interference effect between the CF and DCS amplitudes with $K^0-\overline K^0$ mixing,
\begin{align}\label{po2}
A_{CP}^{\rm int}(t)= -4r_f\cos\phi\sin\delta_f\big(Im(\epsilon)e^{-\Gamma_St}-e^{-\Gamma t}
(Im(\epsilon)\cos(\Delta mt)-Re(\epsilon)\sin(\Delta mt))\big).
 \end{align}

\begin{figure}[tb!]
\centering
\includegraphics[width=.45\textwidth]{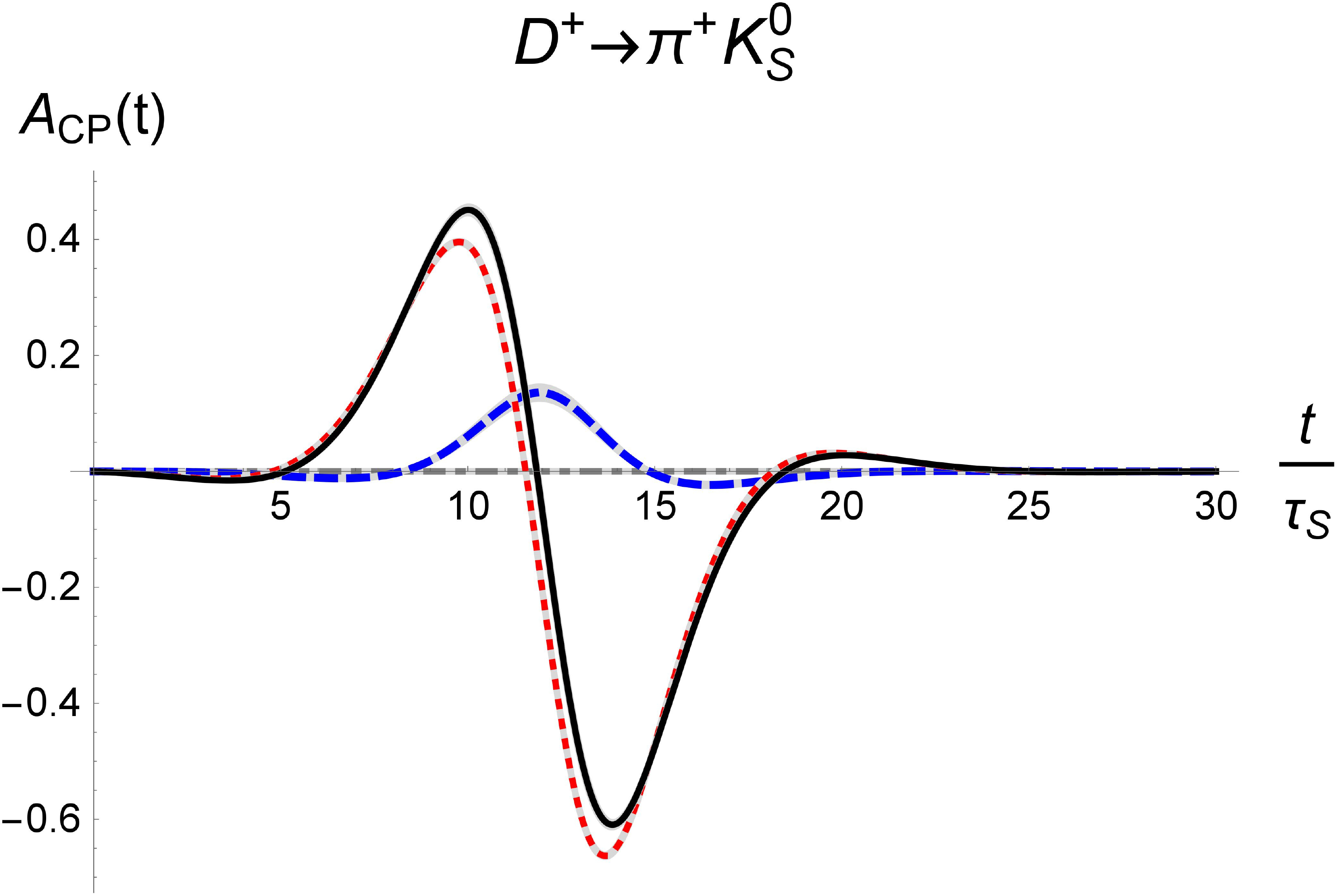}\qquad
\includegraphics[width=.45\textwidth]{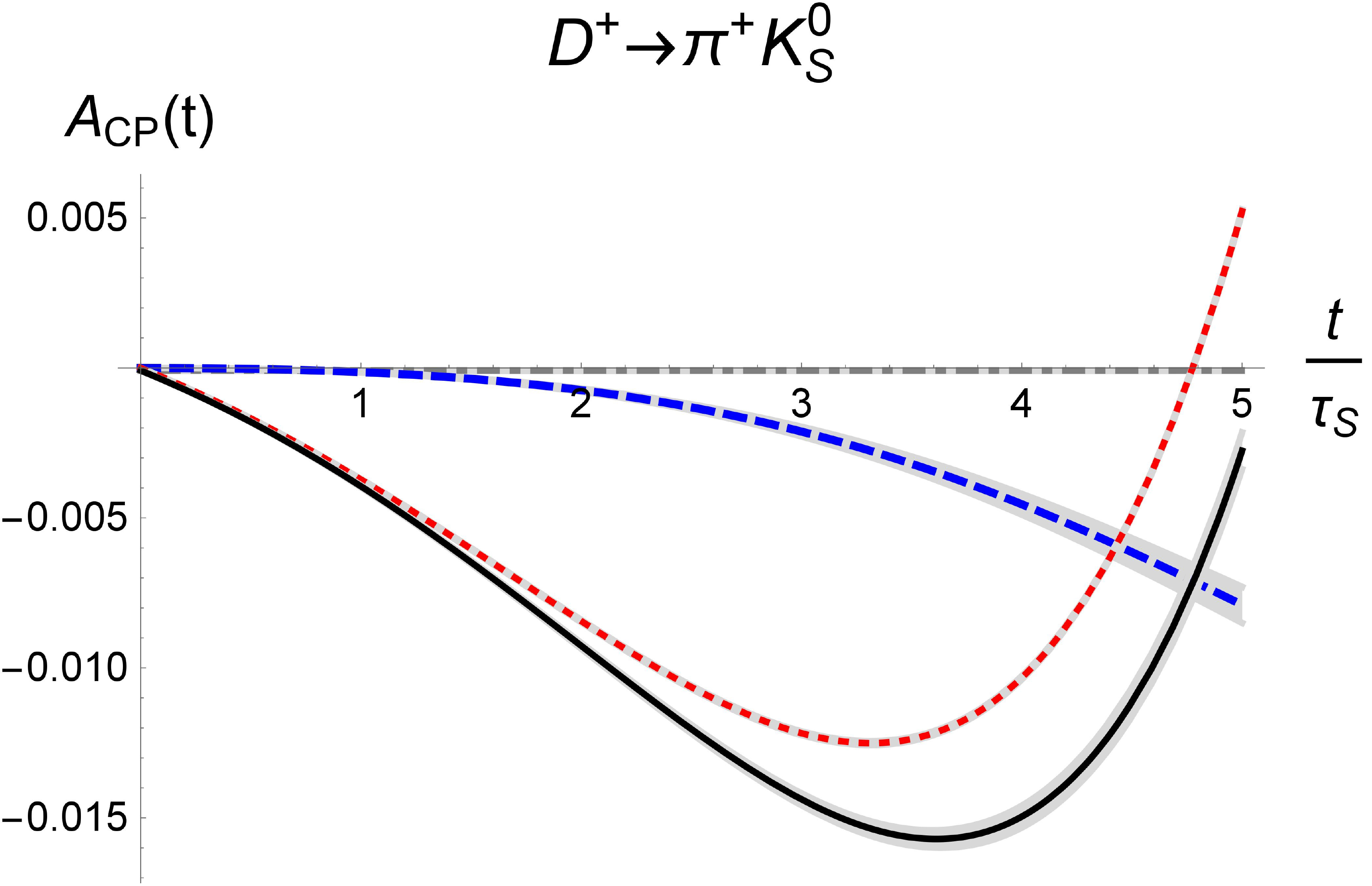}
\includegraphics[width=.4\textwidth]{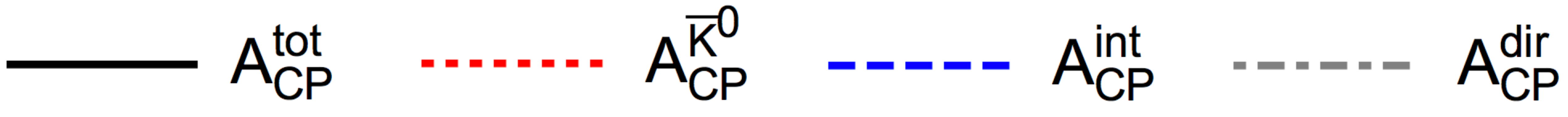}
\caption{The plot of time-dependent $CP$ asymmetries in the $D^+\to \pi^+ K(t)(\to \pi^+\pi^-)$ given by \cite{Yu:2017oky}, where the right figure is the zoomed plot for the small $t$ region of the left one, and the gray bands are the theoretical uncertainties.} \label{fig:ACPt}
\end{figure}

The parameters $r_f$ and $\delta_f$ for the $D^{+}\to \pi^{+}K_{S}^{0}$
and $D_{s}^{+}\to K^{+}K_{S}^{0}$ decays have been estimated in the factorization-assisted
topological-amplitude (FAT) approach \cite{FAT,Wang:2017ksn}.
The dependences of the $CP$ asymmetry in the $D^+\to \pi^+ K(t)(\to \pi^+\pi^-)$
decay on $t/\tau_S$ are displayed in Fig.~\ref{fig:ACPt}.
It is found that the total $CP$ violation dominated by $A_{CP}^{\overline K^0}(t)$, and the deviation from $A_{CP}^{\overline K^0}(t)$ mainly comes from $A_{CP}^{\rm int}(t)$.
The direct $CP$ asymmetries are too small to be seen in Fig.~\ref{fig:ACPt}, being of order of $\mathcal{O}(10^{-5})$.
According to Eqs. \eqref{po1} and \eqref{po2}, $A_{CP}^{\overline K^0}(t=0)=A_{CP}^{\rm int}(t=0)=0$, resulting in $A_{CP}(t=0)=A_{CP}^{\rm dir}(t=0)$.
Both the forthcoming experiments cannot neausre the direct $CP$ asymmetries, unless the large weak phase differences are provided by new physics. Thereby, an observation with nonvanishing
$A_{CP}(t=0)$ indicates new physics.
Compared to the SCS processes, in which the $CP$ asymmetry
cannot discriminate new physics due to the ambiguities in estimating the penguin amplitudes, the
 direct $CP$ asymmetry in neutral kaon modes would give a more unambiguous new physics signal.

\begin{figure}[tph!]
    \centering
\includegraphics[width=.45\textwidth]{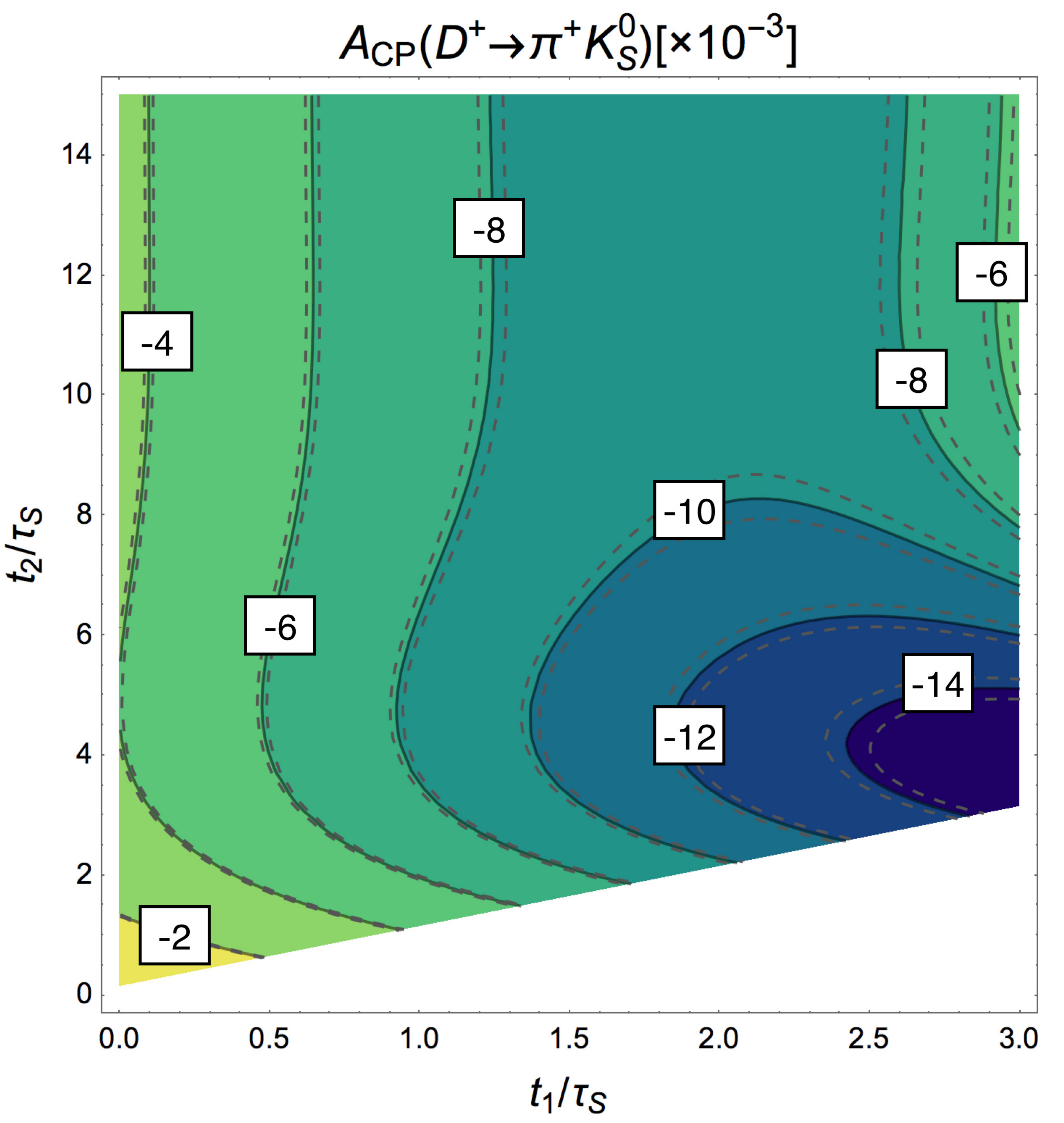}\qquad
\includegraphics[width=.45\textwidth]{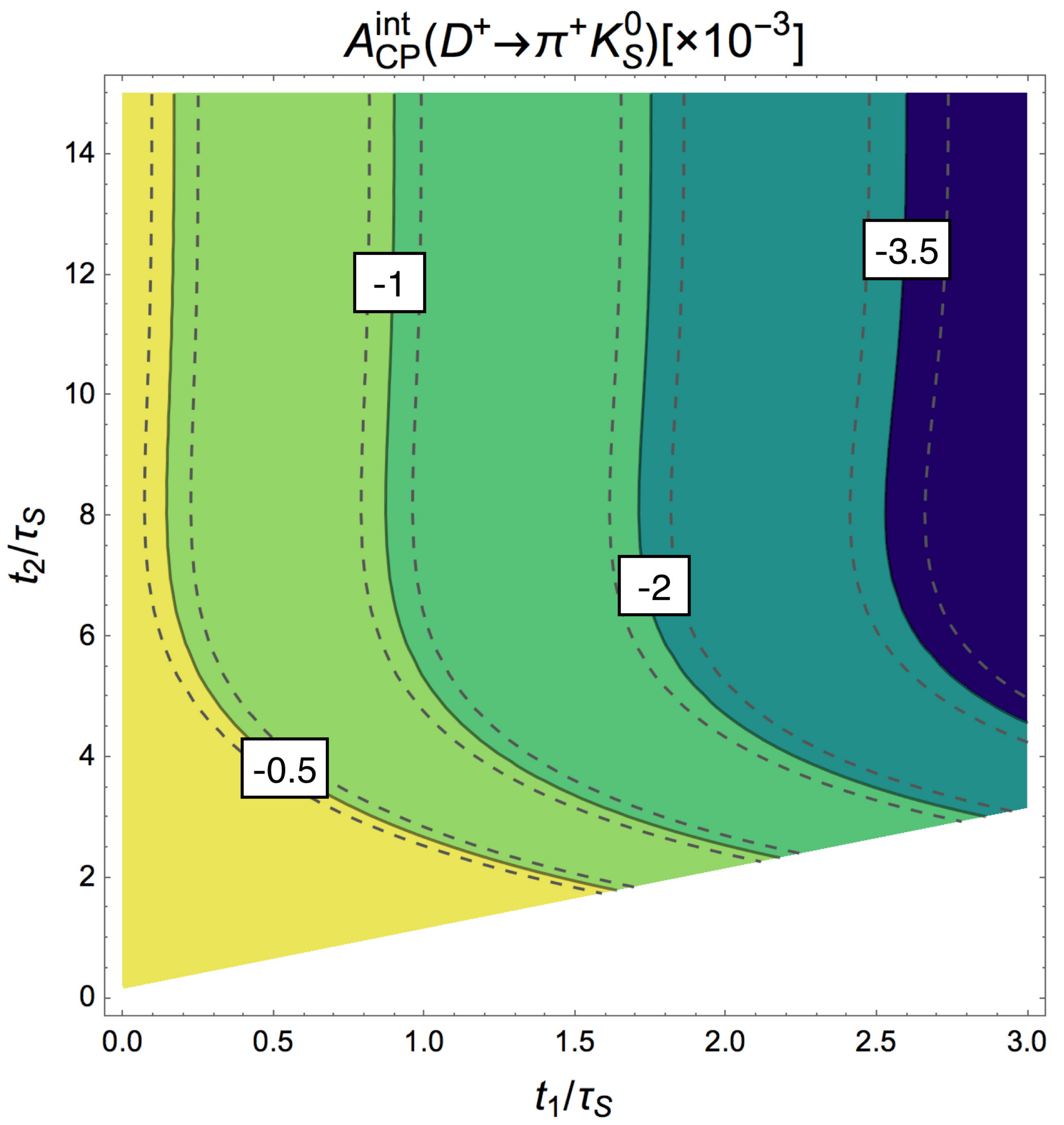}
\caption{The plot of time-integrated $CP$ asymmetries in the $D^+\to \pi^+K_S^0$ given in \cite{Yu:2017oky}, where the left plot is the total $CP$ asymmetry and the right one is the new $CP$-violation effect.
The dashed lines is the theoretical uncertainties of our predictions.
} \label{fig:ACPt1t2}
\end{figure}

The time-integrated $CP$ asymmetry is
\begin{align}\label{time}
A_{CP}=\frac{\int_0^{\infty}F(t)[A_{CP}^{\overline K^0}(t)+A_{CP}^{\rm dir}(t)+A_{CP}^{\rm int}(t)]dt}{\int_0^{\infty}F(t)\,D(t)dt},
\end{align}
where $F(t)$ is a function to take into account
relevant experimental effects. With the approximation of $F(t)=1$ in the interval
$[t_1,t_2]$ and $F(t)=0$ elsewhere \cite{Grossman:2011zk},
quation~(\ref{time}) yields
\begin{eqnarray}\label{eq:Acpt1t2}
 A_{CP}(t_1,t_2)& =\frac{2Re(\epsilon)-4Im(\epsilon)
r_f\cos\phi\sin\delta_f}{1-2r_f\cos\delta_f\cos\phi}\Bigg[1-
\frac{\big[c(t_1)-c(t_2)\big]
            + \frac{Im(\epsilon)+2Re(\epsilon)r_f\cos\phi\sin\delta_f}
                      {Re(\epsilon)-2Im(\epsilon)r_f\cos\phi\sin\delta_f}
               \big[s(t_1)-s(t_2)\big]}
  { \tau_S\Gamma (1+x^2)(e^{-\Gamma_St_1}-e^{-\Gamma_St_2})}
\Bigg]\nonumber\\
&+2r_f\sin\delta_f\sin\phi,
\end{eqnarray}
where $x=\Delta m/\Gamma$, $c(t)=e^{-t \Gamma}[\cos(\Delta m t)-x\, \sin(\Delta m t)]$,
and $s(t)=e^{-t \Gamma}[x \cos(\Delta m t)+ \sin(\Delta m t)]$.
In the first line, those terms proportional to $r_f$ represent the new effect $A^{\rm int}_{CP}(t_1,t_2)$, and those without $r_f$ are the $CP$ violation in the neutral kaon mixing. The second line, which is independent of $t_{1,2}$, corresponds to the direct $CP$ asymmetry in charm decays.
The time-integrated $CP$ asymmetries in the $D^+\to \pi^+K_S^0$ and the new $CP$ violating effect
are exhibited in Fig.~\ref{fig:ACPt1t2}.
In some ranges of $t_1$ and $t_2$, these two quantities are relatively
larger than other ranges.
The experimental investigations could choose the favorable time intervals.
In some experiments, including Belle and LHCb, the new $CP$ violation effect is in absence
\cite{Ko:2012pe,Aaij:2013ula,Aaij:2014gsa,Aaij:2014qec,Aaij:2016dfb}.
However, since this new effect is of the same order as the direct
$CP$ asymmetries in the SCS processes, it cannot be neglected in these measurements.

\begin{figure}[tph!]
\centering
\includegraphics[width=.5\textwidth]{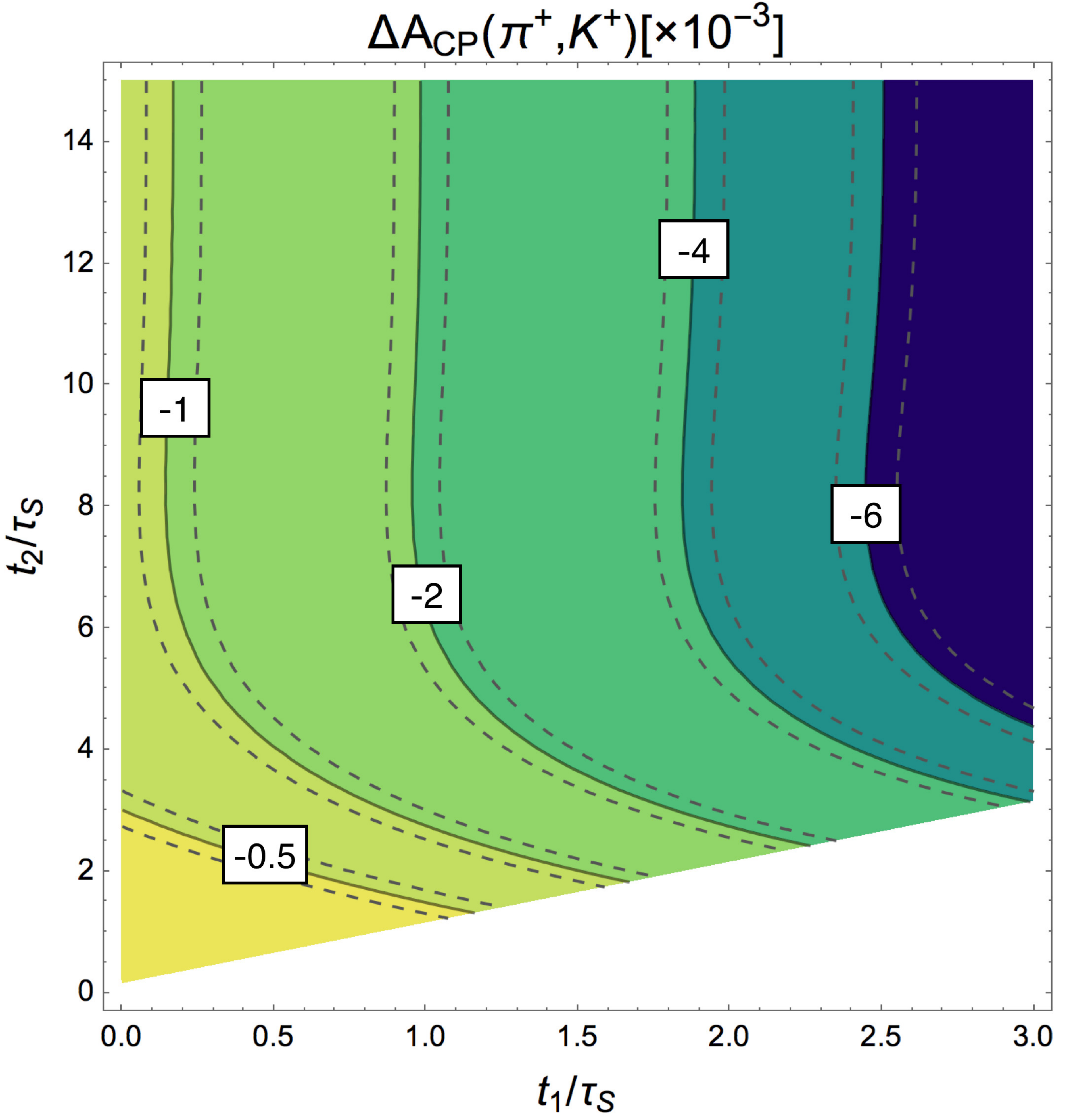}
\caption{The dependences of $\Delta A_{CP}^{\pi^+,K^+}$ on $t_1$ and $t_2$ given in \cite{Yu:2017oky}.}\label{fig:DeltaAcp}
\end{figure}

In order to measure the new $CP$-violation effect in experiments, we propose an
observable
\begin{align}\label{eq:DeltaAcp}
\Delta & A_{CP}^{\pi^+,K^+}\equiv
A_{CP}^{D^+\to \pi^+K_S^0}(t_1,t_2)-A_{CP}^{D^+_s\to K^+K_S^0}(t_1,t_2).
\end{align}
The $CP$ violation in the kaon mixing cancels in the above difference, and
the direct $CP$ violation is negligible.
Our global-fit analysis indicates that the $2r_f\cos\phi\cos\delta_f$ term in denominator of Eq.~\eqref{eq:Acpt1t2} matters little due to the large strong phases $\delta_f$ \cite{Wang:2017ksn}, which is consistent with those derived
in the literature \cite{Muller:2015lua,Bhattacharya:2009ps,Cheng:2010ry,Gao:2014ena} and supported by experiment \cite{He:2007aj}. Then we have
\begin{align}\label{eq:DeltaAcp}
\Delta & A_{CP}^{\pi^+,K^+}
\simeq A_{CP}^{\text{int},D^+\to \pi^+K_S^0}(t_{1},t_{2})
-A_{CP}^{\text{int},D^+_s\to K^+K_S^0}(t_1,t_2).
\end{align}
The model-independent $SU(3)$ symmetry analysis shows the new effects in two modes are constructive
in $\Delta A_{CP}^{\pi^+,K^+}$.
The dependencies of $\Delta A_{CP}^{\pi^+,K^+}$
on $t_1$ and $t_2$ are plotted in Fig.~\ref{fig:DeltaAcp}.
$\Delta A_{CP}^{\pi^+,K^+}$ is of order of $10^{-3}$ in most of time intervals, which is accessible at Belle II and LHCb upgrade experiments \cite{Aaij:2016cfh,Aaij:2014qec,Schwartz:2017gni,Bediaga:2012py}.

In summary, we investigated the time-dependent and time-integrated $CP$ violation in charm decays into neutral kaons.
We first pointed out a new measurable $CP$-violating effect,  the interference between charm decays and kaon mixing, exists in these modes.
It could be revealed by measuring the difference of $CP$ asymmetries in the
$D^+\to \pi^+K_S^0$ and $D_s^+\to K^+K_S^0$ modes on Belle II and LHCb upgrade.
In addition, an observation with non-zero $CP$ violation at $t=0$ would signal new physics.

\vspace*{0.5truecm}

\noindent
{\it Acknowledgements}\\

This work was supported in part by the National Natural Science
Foundation of China under Grants No. 11347027, 11505083, by
the Ministry of Science and Technology of R.O.C. under Grant No.
MOST-104-2112-M-001-037-MY3, and by the Fundamental Research Funds for the
Central Universities under Grant No. lzujbky-2015-241 and lzujbky-2017-97.

\end{document}